\input epsf

\documentclass[12pt]{article}

\usepackage{paper2e}
\usepackage{mydefs2e}

\begin{document}
\begin{titlepage}

\title{Supersymmetric Unification\\ Without
Low Energy Supersymmetry \\ And Signatures for Fine-Tuning at the
LHC}

\author{Nima Arkani-Hamed$^{\rm a}$ and Savas Dimopoulos$^{\rm b}$}

\address{$^{\rm a}$Jefferson Laboratory of Physics, Harvard University\\
Cambridge, Massachusetts 02138}

\address{$^{\rm b}$Physics Department, Stanford University\\
Stanford, California, 94305}

\begin{abstract}

The cosmological constant problem is a failure of naturalness and
suggests that a fine-tuning mechanism is at work, which may also
address the hierarchy problem. An example -- supported by
Weinberg's successful prediction of the cosmological constant --
is the potentially vast landscape of vacua in string theory, where
the existence of galaxies and atoms is promoted to a vacuum
selection criterion. Then, low energy SUSY becomes unnecessary,
and supersymmetry -- if present in the fundamental theory -- can
be broken near the unification scale. All the scalars of the
supersymmetric standard model become ultraheavy, except for a
single finely tuned Higgs. Yet, the fermions of the supersymmetric
standard model can remain light, protected by chiral symmetry, and
account for the successful unification of gauge couplings. This
framework removes all the difficulties of the SSM: the absence of
a light Higgs and sparticles, dimension five proton decay, SUSY
flavor and CP problems, and the cosmological gravitino and moduli
problems. High-scale SUSY breaking raises the mass of the light
Higgs to $\sim 120-150$ GeV. The gluino is strikingly long lived,
and a measurement of its lifetime can determine the ultraheavy
scalar mass scale. Measuring the four Yukawa couplings of the
Higgs to the gauginos and higgsinos precisely tests for high-scale
SUSY. These ideas, if confirmed, will demonstrate that
supersymmetry is present but irrelevant for the hierarchy problem
-- just as it has been irrelevant for the cosmological constant
problem -- strongly suggesting the existence of a fine-tuning
mechanism in nature.

\end{abstract}

\end{titlepage}

\section{Naturalness and its Discontents}
\subsection{Naturalness}
The Standard Model of particle physics is our most successful
physical theory, providing an excellent description of experiments
up to energies of order $\sim 100$ GeV. It is also a consistent
theoretical structure that can be extrapolated by itself up to
energies $\Lambda_{SM}$ far above the weak scale. Yet, ever since
the mid 1970's, there has been a widely held expectation that the
SM must be incomplete already at the $\sim$ TeV scale. The reason
is the principle of naturalness: if $\Lambda_{SM}$ is too large,
the Higgs mass must be fine-tuned to an accuracy of order
$(m_W/\Lambda_{SM})^2$ to explain the weak scale. Solving the
naturalness problem has provided the biggest impetus to
contructing theories of physics beyond the Standard Model, leading
to the proposal of technicolor \cite{Susskind:1978ms} and the
supersymmetric standard model \cite{Dimopoulos:1981zb}, and more
recently, the idea of extra dimensions with low-scale quantum
gravity \cite{Arkani-Hamed:1998nn, Randall:1999ee} and the little
Higgs mechanism \cite{Arkani-Hamed:2001nc}.

Taking naturalness as a principle seriously has had one impressive
concrete success, within the minimal supersymmetric standard model
(SSM): the prediction of gauge coupling unification
\cite{Dimopoulos:1981zb,Dimopoulos:1981yj,Langacker:1995fk} at a
scale $M_G \sim 2 \times 10^{16}$ GeV tantalizingly near the
Planck scale \cite{Dimopoulos:1981yj}. Another success is that
many supersymmetric theories find a good dark matter candidate in
the lightest neutralino \cite{Dimopoulos:1981zb,Goldberg:1983nd}.

Despite these successes, the supersymmetric standard model has
also had a number of difficulties, mostly  having to do with the
fact that the SM explanation for the conservation of baryon,
lepton number and the absence of FCNC's as a consequence of
accidental symmetries disappears in SUSY. There are the well-known
dimension four R-parity violating couplings in the superpotential
that give rise to large proton decay rates and neutrino masses.
Imposing matter or R-parity to forbid these couplings is quite
natural, though, and further ensures the stability of the LSP,
making it a good dark matter candidate. However, there are a
litany of other well-known problems that can not be dispensed with
so elegantly. There are dimension five operators of the form $qq
\tilde{q} \tilde{l}$, that give proton decay. There are new flavor
violations in the the dimension four couplings of fermions to
gauginos and sfermions, that give rise to the SUSY flavor problem.
New CP violating phases have to be significantly suppressed to
avoid large electron and neutron electric dipole moments. There
are also corrections to quantities that do not violate symmetries
any more than in the SM, but which receive significant
contributions from superpartner loops, ranging from $(g - 2)_\mu$
to $B-\bar{B}$ mixing and $b \to s \gamma$. Most important, the
SSM strongly favors a light Higgs, as well as some light
sparticles; their absence is troubling and indicates that there is
already some tuning at the few percent level. Finally, the new
gravitational particles in supersymmetric theories, the gravitino
and moduli, are associated with a variety of cosmological
difficulties.

\begin{figure}
 \centering\leavevmode\epsfysize=8cm \epsfbox{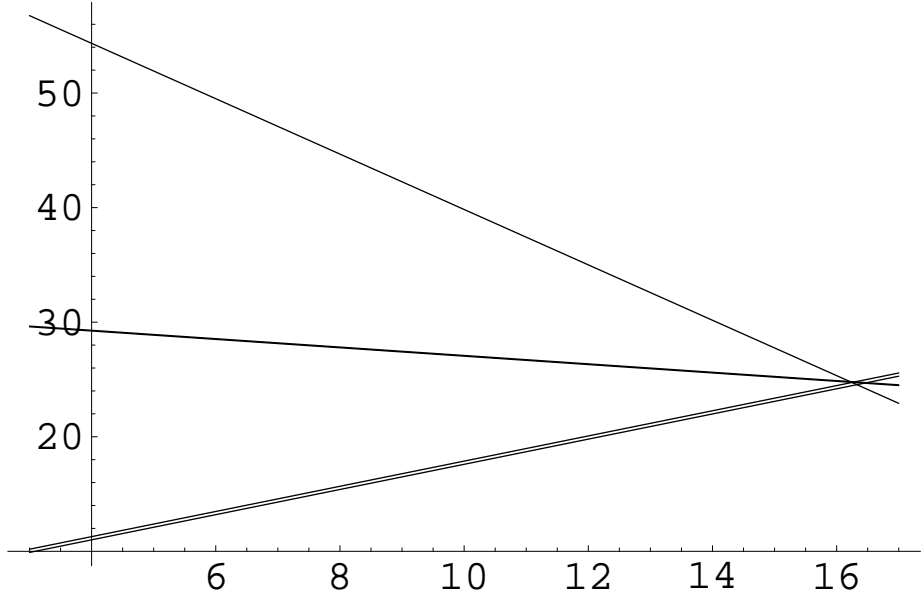}
 \caption{Running couplings in the SSM at one-loop, matching to the full
 SUSY running at $m_t$; from top to bottom,
 $\alpha_1^{-1},\alpha_2^{-1},\alpha_3^{-1}$. We use
 $\alpha_1^{-1}(M_Z) = 58.98 \pm .04, \, \, \alpha_2^{-1}(M_Z) = 29.57
 \pm .03,$ and $\alpha_3^{-1}(M_Z) = 8.40 \pm .14$.}
\end{figure}

\begin{figure}
 \centering\leavevmode\epsfysize=8cm \epsfbox{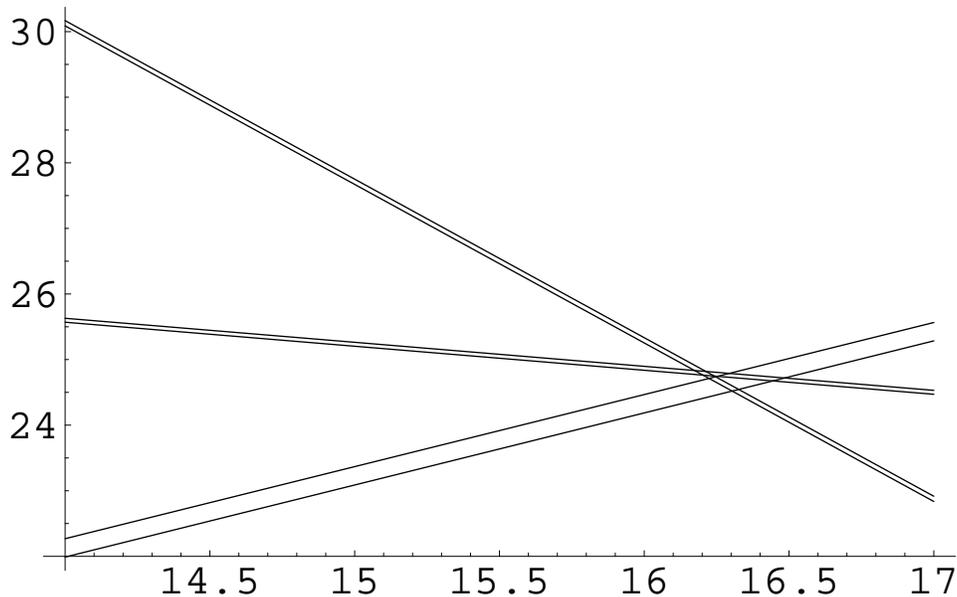}
 \caption{Close-up of the one-loop SSM running couplings near the unification scale.
 Two-loop contributions tend to push up the predicted value of
 $\alpha_3(M_Z)$ to about $.130$, away from the central value $.119$, requiring
 somewhat large compensating threshold corrections.}
\end{figure}

\subsection{A Failure of Naturalness}
Of course none of these challenges are insurmountable, and indeed
attacking them has defined the program of supersymmetric
model-building for the last twenty years. Leaving the basic
structure of the SSM unaltered, various mechanisms have been
invented to address these problems.

In this paper, we will instead suggest a simple but drastic
modification of the usual supersymmetric picture of the world,
which will in a single stroke remove all the phenomenological
difficulties while automatically preserving the concrete successes
of the SSM. In order to motivate our proposal, let us recall the
usual logic leading to the prediction of weak-scale SUSY. Nature
may well be supersymmetric at short distances, perhaps because
SUSY is required for a consistent theory of quantum gravity.
However, given that the low-energy theory does not exhibit
bose-fermi degeneracy, SUSY must be broken. Let the scale of SUSY
breaking in the SSM be $m_S$; the low-energy theory beneath $m_S$
is non-supersymmetric, and therefore the Higgs mass parameter in
this low-energy theory is  UV sensitive. Having $m_h \ll m_S$
would require a fine-tuning, and it would be absurd for the world
to be both supersymmetric {\it and} finely tuned! We therefore
expect that
\begin{equation}
m_h^2 \sim m_S^2
\end{equation}
and so $m_S \lsim$ 1 TeV.

But there is cause to be suspicious of this logic: all of the
theories we study with weak-scale SUSY {\it are} both
supersymmetric and finely tuned, with an enormous fine-tuning for
the cosmological constant. The same line of argument as above
would predict
\begin{equation}
\Lambda \gsim m_S^4
\end{equation}
which is at least 60 orders of magnitude too large.

The usual attitude to the Cosmological Constant Problem has been
one of abhorrence to this fine-tuning, hoping for some deep or
exotic mechanism to explain either why the CC appears so small or
why an enormous vacuum energy doesn't gravitate. Perhaps the CC is
small because of the UV/IR connection, holography and the
mysteries of gravity in deSitter space \cite{Banks:2000fe},
perhaps the graviton is composite at the millimeter scale
\cite{Sundrum:1997js}, or maybe gravity is modified in the IR in a
way that prevents the large vacuum energy from giving rise to an
unacceptable large expansion rate for the universe
\cite{Dvali:2002pe, Arkani-Hamed:2002fu}.

Whatever mechanism may be at work, the fact is that in concrete
theories, the vacuum energy is cancelled by a fine-tuning. For
instance, in supergravity, the positive vacuum energy arising
after supersymmetry breaking is cancelled by adding a constant to
the superpotential. Indeed the gravitino mass arises precisely
from this constant term and therefore is a direct result of the
fine-tuning. {\it Somehow} the enormous $\sim m_S^4$ that we
expect from naturalness must be suppressed
\begin{equation}
\Lambda \sim \epsilon_4 m_S^4
\end{equation}
with $\epsilon_4 \ll 10^{-60}$. Given that this UV sensitive
parameter in the low-energy theory beneath $m_S$ is so much
smaller than its natural size, why are we so confident that the
other UV sensitive parameter, $m_h^2$, must be $\sim m_S^2$?

Again, the usual attitude is that there must be some deep new
physics associated with the CC, since it has to do with gravity,
with all of its associated theoretical mysteries. There doesn't
seem to be anything similarly special about the Higgs mass
parameter. Thus, the philosophy has been to keep the Higgs mass as
natural as possible, while continuing to look for new mechanisms
to solve the cosmological constant problem.

In this paper we wish to explore an orthogonal possibility. What
if the observation of a tiny cosmological constant is telling us
that UV sensitive parameters in the low-energy theory beneath the
SUSY breaking scale will appear incredibly finely tuned? This
leads us to imagine that SUSY is broken in the SSM at very high
scales, far above the weak scale, with the Higgs mass parameter
appearing finely-tuned in the low-energy effective theory, just as
the CC appears finely tuned
\begin{equation}
m_h^2 \sim \epsilon_2 m_S^2
\end{equation}

\subsection{Cosmological Constant Problem and the Emergence of the Landscape}
A possible explanation for such a pattern of fine-tunings can be
found within the context of Weinberg's anthropic resolution of the
CC problem \cite{Weinberg:dv}: if the CC was bigger than about
$\sim 100$ times its observed value, then structure could never
form in our universe; the accelerated expansion due to the CC
would rip apart galaxies before they had a chance to form and the
universe would quickly become empty of everything except the
deSitter Hawking radiation. If there are many different vacua with
different values of the CC, together with a cosmological mechanism
to populate all of them, it is not surprising that we should find
ourselves in a universe with a small enough CC to allow structure
to form, any more than it is surprising that in our own universe
we find ourselves on a tiny planet rather than in the vastly
larger volume of empty space. Note that there is nothing
``anthropic" about this argument, it is really invoking the
``structure" principle (or ``galactic'' principle), the entirely
reasonable statement that we shouldn't expect find ourselves in an
empty universe.

This resolution of the CC problem correctly predicted a small
cosmological constant, and has gained more momentum given that (A)
string theory may well have an enormous ``landscape" of metastable
vacua required to be able to scan the CC finely enough
\cite{Bousso:2000xa, Kachru:2003aw, Maloney:2002rr, Denef:2004ze,
Susskind:2003kw}, and (B) eternal inflation \cite{Linde:eternal}
provides a mechanism by which to populate this landscape. Both of
these ingredients remain controversial \cite{Banks:2003es}. Even
granting these, there are many potential loopholes to the
argument; for instance, if parameters other than the CC vary
significantly in the landscape, then there may be bigger regions
with much larger CC capable of supporting structure. Nevertheless,
it is not implausible that the only parameters that can be
efficiently scanned are the ones that are UV sensitive in the
low-energy theory, and as such can not be controlled by
symmetries.

If the structure principle and the landscape indeed explains the
fine-tuning of the CC, what should we expect for the scale of SUSY
breaking $m_S$? One might think that low-energy SUSY with $m_S
\sim$ TeV is preferred, since this does not entail a large
fine-tuning to keep the Higgs light. However, this conclusion is
unwarranted: the enormity of the CC fine-tuning means that there
are much larger factors in the measure at play. Suppose, for
instance, that we have two regions in the landscape with the
structure of the SSM; in one $m_S$ is $\sim$ TeV and the Higgs
mass is natural, while in the other, $m_S \sim 10^{10}$ GeV and we
have to fine tune by a factor of $\sim 10^{-15}$ for the light
Higgs. But suppose that in the first region there are ``only"
$\sim 10^{40}$ vacua (not enough to be able to find one with a
small enough CC for structure formation), while in second there
are $\sim 10^{200}$ vacua (which {\it is} enough for the tuning of
the CC). Although in the first region the Higgs can be naturally
light without any fine-tuning, there are simply not enough vacua
to find a small enough CC, while in the second region, there are
so many vacua that the additional $\sim 10^{-15}$ tuning to keep
the Higgs light is a small factor in the measure. The point is
clear -- in the landscape picture, the measure is dominated by the
requirement of getting a small enough CC, and since numbers of
order $10^{60}$ are involved, these can dwarf the tuning required
to keep the Higgs light. Without a much better understanding of
the structure of the landscape, we can't decide whether low-energy
SUSY breaking is preferred to SUSY broken at much higher energies.
While it has been argued that low-energy SUSY may be prefered
\cite{Banks:2003es}, this has been questioned in \cite{lenny}.
Furthermore, both the early proposals of \cite{Bousso:2000xa} and
the recent concrete explorations of flux vacua
\cite{Kachru:2003aw, Denef:2004ze} do seem to favor very high
scale SUSY breaking.

If the Higgs mass has to be tuned, there must be some extension of
the ``structure principle" that explains why $m_h^2 \ll m_S^2$. If
in addition to structure we require the existence of atoms, both
Hydrogen and some atom heavier than hydrogen, this ``atomic
principle" can explain the need for the Higgs fine-tuning
\cite{Agrawal:1997gf}. If the Higgs vev decreases by a factor of a
few, the proton becomes heavier than the neutron and Hydrogen
decays. If the vev increases by a factor of a few, the
neutron-proton mass difference becomes far greater the the nuclear
binding energy per nucleon and nuclei heavier than hydrogen decay.
Adopting the ``Carbonic principle", that Carbon must form, gives
an even more precise determination of the Higgs vev. For very
large vevs, the mass difference between the up and down quarks
exceeds the color energy penalty required to have three identical
quarks in a baryon, and the $\Delta^{++}$ becomes the lightest
baryon. The large coulomb barriers and short-range of the strong
interactions prevent the formation of nuclei with multiple
$\Delta^{++}$'s, and the only atoms in the universe would be
chemically identical to Helium. The authors of
\cite{Agrawal:1997gf} performed a systematic analysis of the SM
varying only $m_h^2$, starting from a Higgs vev near the SM value
and going all the way up to $M_{Pl}$, and found that the ``atomic
principle" restricts the Higgs vev to be within about a factor of
$\sim 5$ of its observed value. It is notable that this line of
reasoning also explains one of the striking facts about nature
that is never addressed in conventional theories of physics beyond
the SM: the remarkable proximity of the QCD and electroweak
scales.

With these motivations, we will consider theories in which SUSY is
broken at scales much higher than a TeV, and the fine tuning
required to make the Higgs light happens by some unspecified
mechanism, possibly related to whatever addresses the CCP --
using, for example, the structure and atomic principles as a
selection criterion for the neighborhood of the landscape that we
can find ourselves in.

\section{SUSY without Scalars}

Suppose that SUSY  is broken (in the SSM sector) at a high energy
$m_S$ far above the TeV scale. The scalars of the SSM will then
all be at $m_S$, except for one combination of the two Higgs
doublets that must be finely-tuned to be light. What about the new
{\it fermions} of the SSM, the gauginos and higgsinos? There are
two possibilities: they can also be at the scale $m_S$, or,
because they can be protected by chiral symmetries, they can
survive beneath $m_S$. This is the possibility we wish to pursue.

One reason is that, if these fermions are also near the TeV scale,
gauge coupling unification works essentially identically as in the
SSM. This is because our model differs from the SSM by missing the
squarks and sleptons at low energies, but these scalars come in
complete $SU(5)$ multiplets and do not affect unification at
one-loop order. We are also missing the second Higgs doublet of
the SSM, but this makes a comparatively small contribution to the
beta function, and as we will see, not having it likely {\it
improves} our unification prediction over the SSM when two-loop
corrections are included.

An unrelated reason to expect the gauginos and higgsinos to be
near a TeV, is that this mass scale is independently selected by
requiring the lightest neutralino to be a good dark matter
candidate in our model. Note that, since we are triggering the
weak scale by a fine-tuning, there is no longer a direct link
between Higgs vev and the mass of the gauginos and higgsinos. The
rough link of the dark matter particle mass and the electroweak
vev, which happens naturally in the SSM, is an accident in our
framework. Of course, the accident is not severe; the SM itself is
filled with several  ``accidents" in its spectrum, ranging from
the proximity of the QCD and EW scales to the near equality of the
muon and pion, charm and proton, etc. masses. But as we will see,
in a generic class of models for supersymmetry breaking, we will
in fact {\it predict} the gauginos and Higgsinos to be near the
weak scale, following from the two other high-energy scales
$M_{G}$ and $M_{Pl}$ we know of from a ``see-saw" relation of the
form
\begin{equation}
m_{1/2} \sim \frac{M_{G}^9}{M_{Pl}^8}
\end{equation}
In another class of models, $m_{1/2}$ will be generated by
dimensional transmutation and again come out naturally near the
TeV scale.

We now come to some phenomenological aspects of the low-energy
theory.

\subsection{Finely tuned Higgs}

The most general structure of the low-energy Lagrangian we are
proposing is as follows. All the scalars of the MSSM get
ultraheavy soft masses of order $m_S$. However, one linear
combination of the two Higgs scalars,
\begin{equation}
h = \mathrm{sin} \beta h_u + \mathrm{cos} \beta h_{d}^*
\end{equation}
is fine-tuned to be light.

In more detail, the two Higgs doublets $H_{u,d}$ have soft masses
as well as a $\mu B$ term, so the Higgs boson mass matrix is of
the form
\begin{equation}
\left( \begin{array}{cc} m_u^2 & \mu B \\ \mu B & m_d^2
\end{array} \right)
\end{equation}
(where we have removed a possible phase in $\mu B$ by a field
redefinition). Having a single light Higgs near the weak scale
requires a fine-tuning. The eigenvalues of this mass matrix are
\begin{equation}
\bar{m}^2 \pm \sqrt{\Delta^2  + (\mu B)^2}, \, \, \mbox{where} \,
\, \bar{m}^2 = \frac{m_u^2 + m_d^2}{2},\, \, \Delta =\frac{m_u^2 -
m_d^2}{2}
\end{equation}
 and we require that the smaller of these eigenvalues is
negative but not larger in magnitude than $\sim -m_{EW}^2$.This
requires e.g.
\begin{equation}
(\bar{m}^2)^2 < \Delta^2 + (\mu B)^2 \, \, ,  (\bar{m}^2 +
m_{EW}^2)^2 > \Delta^2 + (\mu B)^2
\end{equation}
We assume $\bar{m}^2$ and $\Delta$ randomly vary over a range
$\sim m_S^2$. As for $\mu B$, it is possible that it too varies
randomly over a range of size $\sim m_S^2$, however, it may be
that since $\mu B$ also further breaks a PQ symmetry on $H_{u,d}$,
it randomly ranges over a range $\sim \epsilon m_S^2$, where
$\epsilon$ is a small parameter characterizing the PQ breaking.

To see the tuning explicitly, let us fix $\mu B$ at $\epsilon
m_S^2$, and randomly vary $\bar{m}^2, \Delta$. The volume of the
region in ($\bar{m}^2,\Delta$) given above, where the light Higgs
is in the tuned range, is then
\begin{equation}
\frac{V_{\mathrm{tuned}}}{V_{\rm{total}}} \sim \frac{m_{EW}^2 \int
d \Delta}{m_S^4} \sim \frac{m_{EW}^2}{m_S^2}
\end{equation}
exhibiting the $\sim (m_{EW}^2/m_S^2)$ tuning needed for the light
Higgs. In this form it is clear that the measure of the tuned
region is dominated by $\Delta \sim m_S^2$. For small $\epsilon$,
getting a light eigenvalue requires $m_u^2 m_d^2 \sim \epsilon^2
m_S^4$, but since the volume of the tuned region is dominated by
$\Delta \sim m_S^2$, in most of region one of $m_{u,d}^2$ is $\sim
m_S^2$ while the other is $\sim \epsilon^2 m_S^4$. If $m_u^2$ is
the small one, the mass matrix has a ``see-saw" form and
tan$\beta$ must be large
\begin{equation}
\mbox{tan} \beta \sim \frac{1}{\epsilon}
\end{equation}
which can help explain the top-bottom mass hierarchy.

There may be natural explanations for why of all the scalars in
the SSM it is only the Higgs that can be light in the low-energy
theory beneath $m_S$. For instance, suppose that the $m^2
\phi^\dagger \phi$ type masses for the scalars stay positive and
$\sim m_S^2$ over the whole range they are scanned. The only
scalars that can even be finely tuned to be light are the ones
that can have $\mu B$-type terms, and in the SSM, these are only
the Higgs doublets.

\subsection{Gauge Coupling Unification as a signal of High-Scale SUSY}

In our model the gauge couplings unify essentially exactly as in
the SSM. Relative to the SSM, we are missing the squarks and
sleptons which come in complete SU(5) multiplets, and therefore do
not affect the unification of couplings at 1-loop. We are also
missing the extra scalar Higgs doublet, which as we will see does
not make a significant contribution to the running.

As we will see later, cosmology favors $m_S$  lighter than $\sim
10^{12} - 10^{13}$ GeV, and in a simple class of models for SUSY
breaking we find $m_S$  near $10^{9}$ GeV. In all cases therefore
some part of the running beneath the GUT scale reverts to the
usual SUSY case. We present the 1-loop evolution of the gauge
couplings for scalars at $10^9$ GeV in Figs. 3 and 4. If as usual
we use the scale where $\alpha_{1,2}^{-1}$ unify to determine the
GUT scale and extrapolate back to predict $\alpha_3(M_Z)$, our
one-loop prediction for $\alpha_3(M_Z) = .108$ is somewhat lower
than in the usual SSM.  This is welcome, because in the SSM, the
two-loop running corrections push up $\alpha_3(M_Z)$ to around
$.130$, somewhat higher than the measured central value of $.119$.
Of course the discrepancy is parametrically within the
uncertainties from GUT scale threshold corrections, although
numerically these have to be somewhat large to compensate for the
discrepancy. While the two-loop corrections in our case are
different than in the SSM and have yet to be calculated, we expect
that they will go in the same direction, pushing our somewhat low
1-loop value for $\alpha_3(M_Z)$ higher, into better agreement
with experiment, requiring smaller compensating threshold
corrections than in the SSM.

\begin{figure}
 \centering\leavevmode\epsfysize=8cm \epsfbox{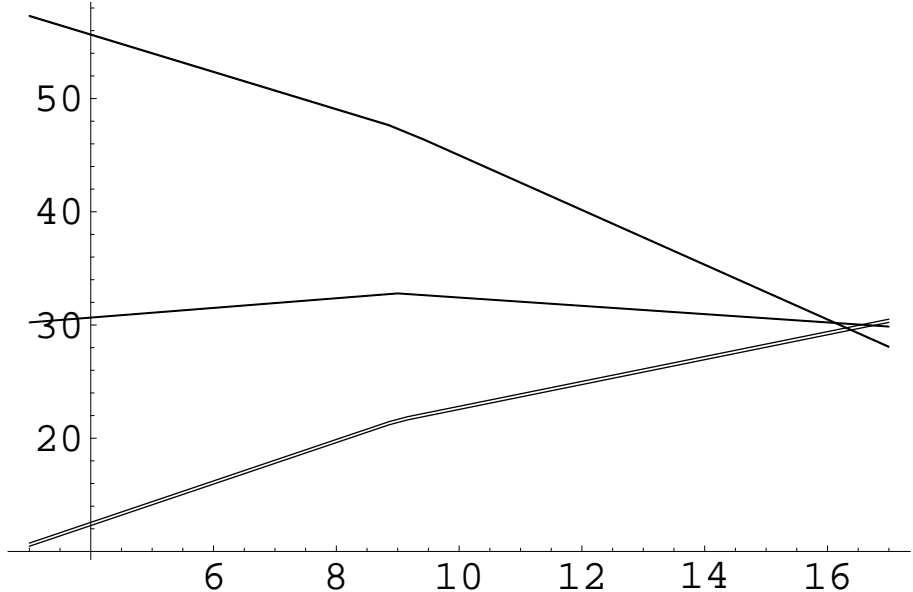}
 \caption{Running couplings in our model at one-loop, with the scalars at $10^{9}$ GeV.}
\end{figure}

\begin{figure}
 \centering\leavevmode\epsfysize=8cm \epsfbox{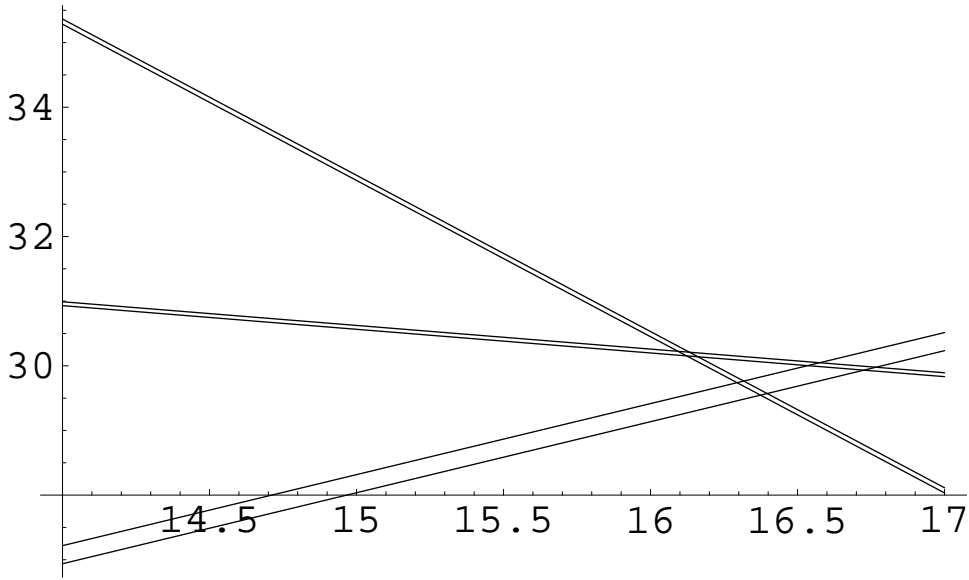}
 \caption{Close-up of the one-loop couplings near the unification scale with the heavy scalars at $10^{9}$
 GeV. Note that the prediction for $\alpha_3(M_Z)$ is lower than
 in the SSM. We expect two-loop corrections to push up
 $\alpha_3(M_Z)$ to better agreement with experiment.}
\end{figure}
\subsection{Effective Lagrangian}

The particle content in the effective theory beneath
$m_S$ consists of the Higgs $h$, as well as the
higgsinos $\psi_{u,d}$, and the gauginos
$\tilde{g},\tilde{b},\tilde{w}$. The most general renormalizable
effective Lagrangian for these fields consists of mass terms for
the fermions, Yukawa couplings between the Higgs and the fermions
and the Higgs quartic coupling:
\begin{eqnarray}
\Delta L & = & M_3 \tilde{g} \tilde{g} + M_2 \tilde{w} \tilde{w} +
M_1 \tilde{b} \tilde{b} + \mu \psi_u \psi_d \nonumber \\
&+& \sqrt{2} \kappa_u h^\dagger \tilde{w} \psi_u +
\sqrt{2}\kappa_d h \tilde{w} \psi_d + \sqrt{2} \frac{1}{2}
\kappa^\prime_u h^\dagger \tilde{b} \psi_u - \sqrt{2} \frac{1}{2}
\kappa^\prime_d
h \tilde{b} \psi_d \nonumber \\
 & - & m^2 h^\dagger h - \frac{\lambda}{2} (h^\dagger h)^2
\end{eqnarray}
Here we have assumed an analog of $R-$ parity, under which all the
new states are odd. As usual, this will ensure that the lightest
of the new fermions is stable and, if it is a neutralino, will be
an good dark matter candidate. Note that even without imposing
R-parity, there are no dimension four baryon number violating
operators in the theory. The reason for imposing R-parity is not
proton decay, but neutrino masses: operators of the form $l h
\tilde{b}$, together with the Majorana mass term for the gauginos,
do violate lepton number, giving rise to unacceptably heavy
neutrinos after EWSB.

\subsection{High scale SUSY boundary conditions and Higgs mass prediction}

At the high scale $m_S$, the four dimensionless couplings
$\kappa_{u,d}, \kappa^\prime_{u,d}$ are determined at tree-level
by the supersymmetric gauge Yukawa couplings of $h_{u,d}$ as

\begin{eqnarray}
\kappa_{u} (m_{S}) &=& g_2(m_{S}) \mbox{sin} \beta , \,
\kappa_{d}(m_{S}) =
g_2(m_{S})\mbox{cos} \beta \\
\kappa^\prime_u (m_{S}) &=&\sqrt{\frac{3}{5}} g_1(m_{S})
\mbox{sin} \beta, \, \kappa^\prime_{d}(m_{S}) = \sqrt{\frac{3}{5}}
 g_1(m_{S}) \mbox{cos} \beta
\end{eqnarray}
where we are using $SU(5)$ normalization for hypercharge. The
Higgs quartic coupling $\lambda$ is determined by the
supersymmetric $D$ terms as usual
\begin{equation}
\lambda(m_{S}) = \frac{\frac{3}{5} g_1^2(m_{S}) + g_2^2(m_{S})}{4}
\, \mbox{cos}^2 2 \beta
\end{equation}
There can of course be threshold corrections to these relations
from integrating out the heavy scalars at the scale $m_S$.

In order to determine the low-energy parameters, we have to run
down from the high scale $m_{S}$ using the RGE's for this
low-energy effective theory. Note that since the theory is not
supersymmetric beneath $m_{S}$, the usual supersymmetric relations
between the Yukawa, quartic and gauge couplings will no longer
hold.

In particular, we will have a significantly different prediction
for the Higgs mass than in the SSM \cite{Okada:1990vk}. Usually in
the SSM, there are two corrections to the Higgs quartic coupling.
First, integrating out the stops generates a threshold correction
to $\lambda$ parametrically of order $(3/8 \pi^2)
(A_t/m_{\tilde{t}})^4$, where $A_t$ is the $A$- parameter
associated with the top Yukawa coupling, which can be a large
correction. Second, there is a log enhanced contribution to
$\lambda$ from the top loop in the low-energy theory beneath
$m_{\tilde{t}}$. Normally with all the scalars near the TeV scale,
these effects are comparable in size, the logarithm is not
particularly big, a full 1-loop analysis is needed, and the Higgs
mass prediction depends on the details of the $A$- terms and stop
spectrum.

The situation is different in our case. First, the same physics
that suppresses the gaugino masses will inevitably also suppress
the $A$ terms so that $A_t \ll m_S$, and the threshold correction
to the Higgs quartic coupling from integrating out the stops at
$m_S$ is tiny, therefore the boundary value for $\lambda(m_S)$ is
accurately given by the tree-level value. Second, with very high
$m_S$, the low-energy Higgs quartic coupling is controlled by the
logarithmically enhanced contribution given by the running the RGE
for $\lambda$ to low energies. This running is quickly dominated
by the contribution from the top Yukawa couplings, and we obtain a
prediction for the Higgs mass.

At 1-loop, the RGE for $\lambda$ in the theory beneath $m_S$ is
\begin{eqnarray}
16 \pi^2 \frac{d \lambda}{d t} &=& 12 \lambda^2 + \lambda \left(12
\lambda_t^2 + 6 \kappa_u^2 + 6 \kappa_d^2 + 2 \kappa_u^{\prime 2}
+ 2 \kappa_d^{\prime 2} \right) \nonumber \\ & - & 3 \lambda
\left( 3 g_2^2 + \frac{3}{5} g_1^2 \right)+ \frac{3}{4}(2 g_2^4 +
(g_2^2 +
\frac{3}{5} g_1^2)^2)  \nonumber \\
&-& \left(12 \lambda_t^4 + 5 \kappa_{u}^4 + 5 \kappa_{d}^4 +
\kappa_u^{\prime 4} + \kappa_d^{\prime 4}\right) \nonumber \\ &-&
\left(2 \kappa_{u}^2 \kappa_{u}^{\prime 2} + 2 \kappa_d^2
\kappa_d^{\prime 2} + 2 \kappa_d^2 \kappa_u^2 + 2 \kappa_u^{\prime
2} \kappa_d^{\prime 2} + 4 \kappa_u \kappa_u^\prime \kappa_d
\kappa_d^\prime \right)
\end{eqnarray}

A complete analysis of the Higgs mass prediction at one loop would
require solving the coupled RGE's for $\lambda$ together with the
top Yukawa coupling $\lambda_t$ and the $\kappa$'s. But the
largest contribution to $\lambda$ come from the top Yukawa and are
$\propto \lambda_t^4$, so the Higgs mass depends most sensitively
on the top Yukawa coupling.  Extracting $\lambda_t$ from the top
mass at tree-level gives $\lambda_t = 1$, however, when the 1-loop
QCD corrections to the top mass are taken into account, this is
reduced to $\lambda_t = .95$, decreasing the Higgs mass prediction
by $\sim 20$ \%. Meanwhile, the recent CDF/D0 measurements of the
top mass point to a somewhat heavier top quark $\sim 178$ GeV
\cite{Group:2004rc}. These uncertainties involving the top Yukawa
have a bigger impact on our Higgs mass prediction than all the
terms involving the $\kappa$ and $g_{2,1}$ couplings in the
$\lambda$ RGE's. Nonetheless as a preliminary analysis that we
believe will capture the most important effects, we numerically
solve the 1-loop RGE's for $\lambda, \lambda_t$ and the
$\kappa$'s, using the boundary condition $\lambda_t(m_t) = .95$
for $m_t = 174$ GeV and $\lambda_t(m_t) = .97$ for $m_t = 178$
GeV, while keeping only the contributions from the largest
couplings $\lambda_t,g_3$ in the $\lambda_t,\kappa$ runnings
\begin{eqnarray}
16 \pi^2 \frac{d \lambda_t}{dt} &=& \lambda_t (\frac{9}{2}
\lambda_t^2 - 8 g_3^2) + \cdots \\ 16 \pi^2 \frac{d
\kappa^{(\prime)}_{u,d}}{d t}
 &=& 3 \lambda_t^2 \kappa^{(\prime)}_{u,d} + \cdots
\end{eqnarray}
and using the high-scale SUSY boundary conditions for all the
couplings. The prediction for the low-energy Higgs mass
\begin{equation} m_h \sim \sqrt{\lambda} v
\end{equation}
with $\lambda$ evaluated at $\sim m_t$, is plotted in Fig. 5, as a
function of the the scale $m_{S}$ and tan$\beta$. We find $m_h$ in
the range between $\sim$ 120-150 GeV, light but above the LEPII
limits.

\begin{figure}
 \centering\leavevmode\epsfysize=8cm \epsfbox{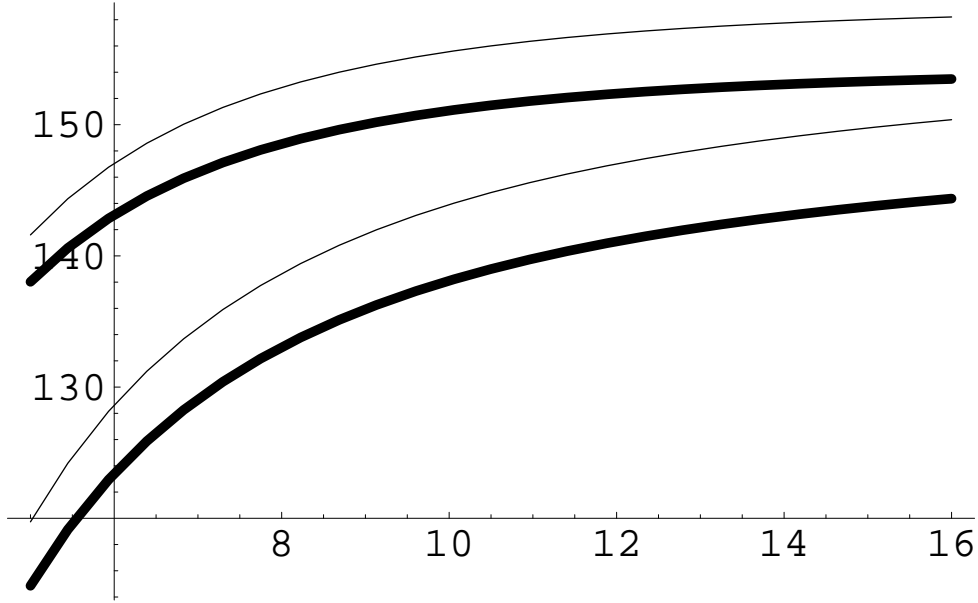}
 \caption{The Higgs mass in GeV, as a function of log$_{10} (m_{S}/\mbox{GeV}$).
The thick line is for $m_t = 174$ GeV, the thin line for $m_t =
178$ GeV, while lower lines are for cos$2 \beta = 0$ and the upper
lines for cos$2 \beta = 1$.}
\end{figure}

The Higgs mass can give a first, very rough estimate of the scale
$m_S$. But in principle, all the couplings $\lambda, \kappa_{u,d},
\kappa^\prime_{u,d}$ can be measured at low energies, and running
them to high energies should show that all five of them hit their
supersymmetric values at the same scale $m_S$. A convenient
digramatic representation of this is to group $\kappa_{u,d}$ into
a two-dimensional vector $\vec{\kappa} = \kappa_d \hat{x} +
\kappa_u \hat{y}$, and $\kappa^\prime_{u,d}$ into the vector
$\sqrt{3/5} \vec{\kappa}^\prime = \kappa^\prime_d \hat{x} +
\kappa^\prime_u \hat{y}$. Running these vectors to high energies,
the SUSY boundary conditions tell us that at the scale $m_S$ these
two vectors must be aligned in the same direction, with angle from
the horizontal $\beta$, and that the lengths of $\vec{\kappa},
\vec{\kappa^\prime}$ should be $g_2(m_S), g_1(m_S)$ respectively.
Having determined $\beta$, one can also check that the running
Higgs quartic $\lambda(m_S)$ hits its supersymmetric value. These
checks are illustrated in Fig. 6. Clearly if all of these
measurements were made and  these predictions confirmed, it would
be striking quantitative evidence for {\it high scale}
supersymmetry at a scale $m_S$, with a finely tuned Higgs in the
theory beneath $m_S$.

\begin{figure}
 \centering\leavevmode\epsfysize=8cm \epsfbox{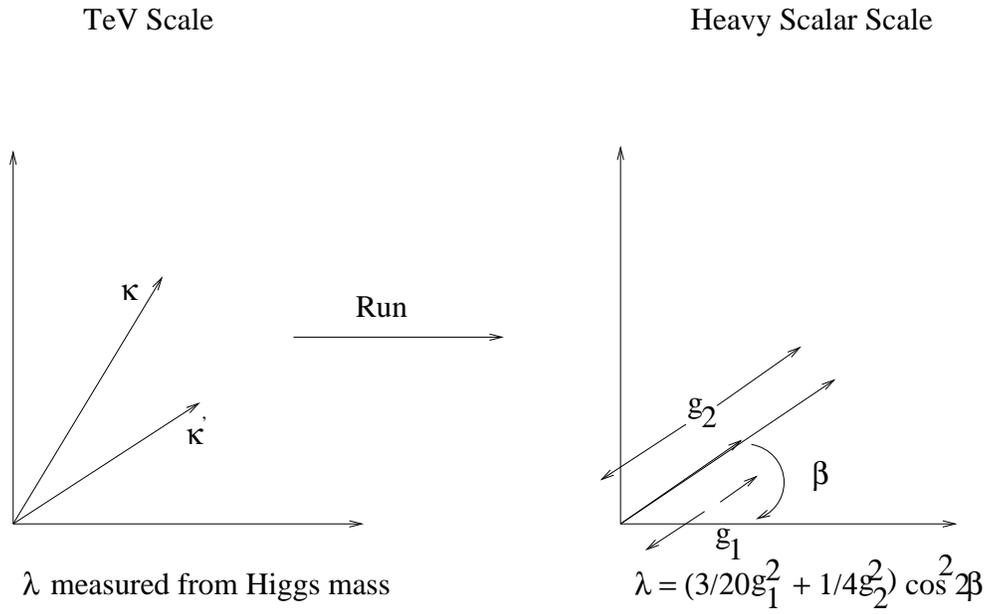}
 \caption{Evidence for high scale SUSY from running the couplings $\kappa_{u,d},
 \kappa^\prime_{u,d}$, grouped into 2-d vectors, from low to high energies. The two vectors align at a scale
 $m_S$ where they must have lengths $g_{2,1}$. This must agree with the $m_S$ extracted from the gluino lifetime.
 $\beta$ is
 determined, and fixes $\lambda(m_s)$, which can be checked against the $\lambda$ determined by the Higgs mass.}
\end{figure}

We have discussed the Higgs mass and $\kappa,\kappa^\prime$
predictions in our minimal model, but these can change in less
minimal models by a number of new factors absent in the usual SSM.
Since the evolution of the couplings beneath $m_S$ is
non-supersymmetric and the supersymmetric link between $\lambda,
\kappa_{u,d}, \kappa_{u,d}^\prime$ and the gauge couplings
$g_{1,2}$ no longer holds, the presence of additional vector-like
matter multiplets (say a number of $(5 + \bar{5})$'s) in the
theory beneath $m_S$ can affect the Higgs mass prediction. There
may also be new contributions to the Higgs quartic coupling at
$m_S$, coming from additional superpotential or $D$-term
couplings. These are now only constrained by the requirement of
perturbativity from $m_S$ to $M_G$, and may therefore give larger
corrections to the Higgs mass than in the usual SSM. These
interesting issues deserve further exploration.

\subsection{Long-Lived Gluino as a probe of Fine-Tuning}
A striking qualitative prediction of our new framework, decisively
differentiating it from the usual SSM, is the longevity of the
gluino. Because the scale of supersymmetry breaking is now high,
the squarks are heavy and the lifetime for the gluino to decay
into a quark, antiquark and LSP -- which is mediated by virtual
squark exchange -- becomes very long, of order
\begin{equation}
\tau = 3 \times 10^{-2} \rm{sec} \Big(\frac{m_S}{10^9\,
GeV}\Big)^4 \Big( \frac{1\, \rm{TeV}}{m_{\tilde{g}}}\Big)^5,
\end{equation}
where $m_S$ is the squark mass, $m_{\tilde{g}}$ the gluino mass.
We have included a QCD enhancement factor of $\sim 10$ in the
rate, as well as another factor $\sim 10$ for the number of decay
channels. The gluino lifetime can easily range from $10^{-6}$ sec
to the age of the universe, as  $m_{S}$ ranges from $10^8$ GeV to
$ 10^{13}$ GeV, and $m_{\tilde{g}}$ from $100$ GeV to 1 TeV. As
long as its lifetime is much longer than $10^{-6}$ sec, a
$\it{typical}$ gluino produced at the LHC will decay far outside
the detector. This is a key difference between our theories and
the SSM, fundamentally tied to the large-scale breaking of
supersymmetry, and, once the gluino is produced at the LHC, can
immediately experimentally distinguish our model from a
conventional hierarchy-motivated SUSY theory with scalars just
barely too heavy to be produced (say at $\sim 10$ TeV), where the
gluino would still decay well inside the detector.

The only trace of a  typical gluino decaying outside the detector
will be the energy that it deposits in the detector
\cite{Baer:1998pg,Raby:1998xr,Culbertson:2000am}. However, at peak
luminosity of 30 fb$^{-1}$ per year, the LHC may well be a gluino
factory producing roughly a gluino per second (for $m_{\tilde{g}}
\sim 300$ GeV). It is therefore possible to get statistically
important information by relying on \emph{atypical} events involving:\\
\textbf{Displaced gluinos:} These are simply gluinos which decay
in the detector, even though their lifetime is longer than the
size of the detector. The number of these events will become too
small once the lifetime becomes longer than roughly one second.
For $m_S \sim 1000$ TeV or so,  all the gluinos will decay inside
the detector, but may live long enough to have displaced vertices.\\
\textbf{Stopped, long-lived gluinos:} These are gluinos which lose
energy and stop in the detector or in the surrounding earth, and
decay much later -- seconds, days, or months later, perhaps even
when there is no beam in the accelerator! The lifetime sensitivity
can extend to $\sim 10^7$ years, corresponding to a SUSY breaking
scale $m_S $ of up to $\sim 10^{13}$ GeV, and depends on the
fraction of gluinos that stop in the detector; this is involved
and in turn depends on the fraction of time the gluino dresses
into  a charged hadron and loses energy electromagnetically. Such
events can be spectacular and give charged tracks, displaced
vertices, delayed decays, and possibly even intermittent tracks,
all at the same time. It is also possible to just have displaced
vertices, and delayed decays, without charged tracks. Since the
final decays can occur much after the collision that created the
gluino, triggering on these poses interesting challenges. For long
lifetimes, a good time to look for such events is when there is no
beam, but the detectors are on. A particularly good place to look
is at the endpoints of charged tracks.

\subsection{Gluino Cosmology}

In our framework, there are two particles that are potentially
important for cosmology. One is the  LSP neutralino, a natural
candidate for the DM particle \cite{Dimopoulos:1981zb}, as in most
versions of the SSM. Another is the gluino, which is now long
lived.

Gluinos can be cosmologically excluded either because their
abundance today is unacceptably large or, if their lifetime is
shorter than the age of the universe, their decay products can
distort the photon background or destroy nuclei synthesized during
primordial nucleosynthesis, which began when the universe was one
second old. A gluino that decays in less than a second is
harmless, as its decay products thermalize
and the heat bath erases any trace of its existence.
Gluinos that live longer than a second can be safe, as long as
their abundance is small.

We now turn to an estimate of the abundance of gluinos before they
decay. When the temperature of the universe drops below
$m_{\tilde{g}}$, the gluino's abundance is maintained in thermal
equilibrium by their annihilation into gluons. Eventually, their
abundance becomes so low that they cannot find each other in the
expanding universe, they stop annihilating, and their abundance
``freezes out". There are three stages of gluino annihilation,
characterized by three different processes  by which gluinos can
annihilate. In chronological order, they are: perturbative
annihilation, annihilation via recombination, and annihilation in
the QCD era. The first ends at a
 temperature of order $T_F \sim 1/27\, m_{\tilde{g}}$ and 
, as long as $T_F \gg \Lambda_{QCD}$, 
leads to the  canonical fractional abundance today of order,
\begin{equation}
\frac {\rho_{\tilde{g}}}{\rho_T} = \frac {m_{\tilde{g}}^2}{N\,
\alpha_s^2 \, (100\,\mbox{TeV})^2}
\end{equation}
where $N$ is a numerical factor, depending on the number of decay
channels, color factor etc., which we estimate to be of order 100.
Next, when the universe cools down  to a temperature below the
gluinium (a bound state of two gluinos) binding energy $E_B$,
\begin{equation}
T \lsim E_B = \frac{1}{2}\, m_{\tilde{g}}\,\alpha_s^2 \sim 5\,
\mbox{GeV} \Big( \frac{m_{\tilde{g}}}{1 \mbox{TeV}} \Big)
\end{equation}
a new period of possible gluino annihilation begins. Gluino pairs
can now ``recombine'' (via gluon emission) into a gluinium  which
is no longer broken apart by the ambient  thermal gluons. The
recombination cross section, however, is comparable to the
perturbative annihilation cross section, so no significant further
gluino annihilation occurs.

Next, the temperature drops to $\Lambda_{QCD}$ and the strongly
interacting particles organize themselves into a dilute gas of
color-singlet baryons, R-hadrons and gluinia. What happens beyond
this is difficult to analyze quantitatively, as it involves hadron
dynamics \cite{Baer:1998pg,Raby:1998xr}.

One scenario that we consider plausible is that when two slow
R-hadrons collide they recombine into a bound R-molecule (by
emitting a pion), containing two gluino ``nuclei", with a cross
section of order $\sigma \sim 30$ mb. Subsequently, the two
gluinos inside this small molecule rapidly find and annihilate
each other into gluons, before the molecule has a chance to be
dissociated by collisions with the dilute gas particles. This
avoids the suppression $\sim m_{\tilde{g}}^{-2}$ in the
perturbative annihilation cross-section, and results in a small
gluino abundance which we estimate by equating expansion and
reaction rates,$
 n\, \sigma\, v \sim T^2/M_{Pl}$
where $T \sim \Lambda_{QCD}$. This translates to
\begin{equation}
\frac {n_{\tilde{g}}}{n_{\gamma}} = 10^{-18} \left(\frac
{m_{\tilde{g}}}{ 1\, \mbox{TeV}}\right)^{1/2}
\label{eq:abundance1}
\end{equation}
\begin{equation}
m_{\tilde{g}}\, \frac {n_{\tilde{g}}}{n_{\gamma}} = 10^{-15}
\left(\frac {m_{\tilde{g}}}{ 1\,\mbox{TeV}}\right)^{3/2} \,
\mbox{GeV} \label{eq:abundance2}
\end{equation}
The last quantity measures the destructive power of the decaying
gluino gas, as it depends on both the mass and the concentration
of gluinos. The abundance of gluinos with lifetimes comparable to
the age of the universe is constrained by the the negative
searches for abnormally heavy hydrogen, helium, and lithium that
would have formed during primordial nucleoynthesis, as well as the
limits on stable strongly interacting massive
particles~\cite{Starkman:1990nj} that would result from the
pairing of a gluino and a gluon. These limits are much stronger
than the abundance of equation(~\ref{eq:abundance1}), in the case
of heavy hydrogen by a factor of $10^{22}$  \cite{Smith:qu}. So,
the gluino lifetime must be shorter than roughly $10^{16}$
seconds, corresponding to an upper limit of about $m_S \lsim 3
\times 10^{13}$ GeV, for a 1 TeV gluino. Evading this cosmological
limit on $m_S$ is possible in theories where the reheat
temperature is much lower than the gluino mass, so that gluinos
are not produced after reheating.

The abundance of gluinos with lifetime up to $10^{13}$ sec must be
small to avoid  spectral distortions of the CMBR \cite{Hu:gc}.
This constraint is mild, and equation (\ref{eq:abundance2}) easily
satisfies it. The abundance of gluinos with lifetime in the range
from $10^{-1}$ sec to $10^{12}$ sec must also be small to avoid
the destruction of the light nuclei synthesized during the
BBN~\cite{Dimopoulos:1988ue,Kawasaki:2004yh}. Although this
constraint is strong, especially for lifetimes between $10^{4}$
sec to $10^{7}$ sec, equation (\ref{eq:abundance2}) satisfies it.
Other constraints from possible distortions of the diffuse photon
background  are easily satisfied.

The problem of computing the gluino abundance through the QCD era
is  important and should be revisited~\cite{Baer:1998pg}. We
stress that our picture for gluino annihilation after the QCD
phase transition is rough and may be missing important effects
that suppresses the annihilation and increases the abundance,
which may lead to better limits for the gluino mass and the scale
of SUSY breaking in our framework.

\subsection{Addressing the Problems of the SSM}

The SSM has many phenomenological problems associated with the 110
independent parameters in the flavor sector alone
\cite{Dimopoulos:1995ju}. These problems originate in the 97
parameters that reside in the scalar sector, in the mass- and
A-matrices of squarks and sleptons (the rest are just the usual KM
parameters); it is hard to hide all 97 parameters of the flavor
sector of the SSM from low energy physics and avoid problems with
a large number of rare processes such as, FCNCs, CP-violation,
b-decays -- when the scalars are near a TeV, as required by the
naturalness. Starting with the original universality hypothesis
\cite{Dimopoulos:1981zb}, much of the model building in the last
23 years has been targeted to solving these flavor problems by
attempting to derive universality from some specific dynamics --
such as gravity \cite{Barbieri:1982eh}, gauge \cite{Dine:1981gu,
Dine:1995ag}, anomaly \cite{Randall:1998uk,anom} and gaugino
\cite{Chacko:1999mi} mediation -- in spite of the violations of
flavor in the Yukawa couplings of quarks. In addition there are
difficulties associated with dimension five proton decay operators
and CP violating SUSY phases. Meanwhile, the absence of a light
Higgs at LEPII has raised new problems, necessitating tunings at
the few percent level for electroweak symmetry breaking
\cite{Barbieri:2000gf}.

All these problems evaporate as soon as we raise the scale of
sparticles to $\sim 100-1000$ TeV. The physical relevance of all
the 97 parameters connected to the flavor problem disappears
because they are linked to the scalars that now decouple.
Similarly for the proton lifetime via the dimension 5 operators.
As we have seen the light Higgs is naturally heavier; of course
there is tuning to get the Higgs mass light, but unlike the usual
SSM, naturalness is not our guiding principle, and we have argued
that this tuning is taken care of by the ``atomic principle".
Finally, while there may be phases in $\mu$ and the $M_i$, these
first affect only the Higgs sector at 1-loop, and only much more
indirectly feed into the electron and neutron edms, which are
naturally small enough.

In addition the SSM has problems of cosmological origin, the
gravitino and moduli problems. As soon as the scale of SUSY
breaking is raised to over $\sim 100$ TeV, the gravitino and
moduli decay with lifetimes less than a second, and these problems
also evaporate.

\section{Models}

We now give examples of models where there is a natural separation
of scales between the scalar and the gaugino/higgsino masses, with
chiral symmetries keeping the fermions light relative to the
scalars. We will begin by considering very standard sorts of
models where the low-energy theory beneath the cutoff contains
supergravity. In such a theory, the only way to cancel the vacuum
energy after supersymmetry breaking is to add a constant $c$ to
the superpotential, which breaks R-symmetry and makes the
gravitino massive. Since $R$ is neccessarily broken, at some level
we must induce a gaugino mass.

\subsection{Anomaly mediated gaugino masses with scalars at $\sim 1000$ TeV}

If we assume that the field $Z$ breaking SUSY ($F_Z \neq 0$)
carries some symmetry so that the operator $\int d^2 \theta Z WW$
is forbidden, then the leading source for a gaugino mass is from
anomaly mediation \cite{Randall:1998uk,anom}. Since $R$ is broken,
the $F$ component of the chiral compensator field $\phi$ in
supergravity can be non-zero, yielding gaugino masses of order
\begin{equation}
m_{1/2} \sim \frac{g^2}{16 \pi^2} F_\phi
\end{equation}
If $F_\phi$ is $\sim m_{3/2}$, this limits $m_{3/2} \lsim 50$ TeV
for gauginos near the TeV scale. In \cite{anom} and more recently
in \cite{Wells:2003tf}, the phenomenology of scalars with mass
$\sim m_{3/2}$ has been explored. From our point of view, however,
this is not heavy enough -- scalars at $\sim 50$ TeV are in an
uncomfortable no-man's land between being natural and tuned, and
do not in themselves solve e.g. the SUSY flavor problem outright.
Fortunately, the scalars can be much heavier than $m_{3/2}$, since
the operators in the K\"{a}hler potential giving the scalar masses
can be suppressed by a fundamental scale $M_* \sim M_G$ much
smaller than the Planck scale, so that
\begin{equation}
m_{3/2} \sim \frac{F_Z}{M_{Pl}}, \sim 50 \mbox{TeV}, \, \, m_S
\sim \frac{F_Z}{M_*} \sim 500 - 5000 \, \mbox{TeV}.
\end{equation}
For $\mu$ and $\mu B$, we can simply write down the usual
Giudice-Masiero \cite{Giudice:1988yz} operators
\begin{equation}
\int d^4 \theta \frac{Z^\dagger}{M_*} H_{u} H_d, \, \, \int d^4
\theta \frac{Z^\dagger Z}{M_*^2} H_u H_d
\end{equation}
If both of the above operators have an additional suppression by a
factor of $\epsilon$ since they break a PQ symmetry, then we have
\begin{equation}
(\mu B) \sim \epsilon (100 \mbox{TeV})^2, \mu \sim \epsilon (100
\mbox {TeV})
\end{equation}
Recall that small $\epsilon$ leads to large tan$\beta \sim
1/\epsilon$, which is natural to be $\sim m_t/m_b$; in this case
we can find $\mu$ close to the TeV scale
\begin{equation}
\mu \sim \frac{m_b}{m_t} \times 100 \mbox{TeV} \sim \mbox{TeV}
\end{equation}

This set-up is very generic, and it can push scalars up to masses
$\gsim$ 1000 TeV, high enough to evade all phenomenological
problems. This is also in an interesting range for gluino collider
phenomenology: all the gluinos can decay inside the detector, but
with a long enough lifetime to have observable displaced vertices.
Together with the gaugino masses, this is a smoking gun for
anomaly mediation with ultraheavy scalars. In addition, as usual
in anomaly mediation, with $m_{3/2} \sim 50$ TeV the gravitino and
moduli problems disappear.

\subsection{Theories with $\sim 100$ GeV Gauginos and Higgsinos: Generalities}

It is possible to consider a large class of theories where the
$F_{\phi} \ll m_{3/2}$, and the anomaly mediated gaugino masses
are negligible relative to other sources of SUSY breaking. The
leading $R$-invariant operator we can write down that generates a
gaugino mass directly from $R$ and SUSY breaking is
\begin{equation}
\int d^4 \theta \frac{Z^\dagger Z c^\dagger}{M_{Pl}^5} W_\alpha
W^\alpha
\end{equation}
Similarly, the leading $R$- invariant operator leading to a
non-zero $\mu$ (given $H_{u,d}$ have R-charge 0) is
\begin{equation}
\int d^4 \theta \frac{Z^\dagger Z c}{M_{Pl}^5} D_\alpha H_u
D_\alpha H_d
\end{equation}
Recalling that the gravitino mass is $m_{3/2} \sim |c|/M_{Pl}^2$,
and that we must have $3 |c|^2/M_{Pl}^2 = F_X^2$ to cancel the
vacuum energy, leads to a gaugino/Higgsino mass
\begin{equation}
m_{1/2} \sim \mu \sim \frac{m_{3/2}^3}{M_{Pl}^2}
\end{equation}
If we assume that $c \sim M_{G}^3$, the most natural value in a
theory where the fundamental scale is near $M_G$, we find
\begin{equation}
m_{3/2} \sim \frac{M_G^3}{M_{Pl}^2}, \sim 10^{13} \mbox{GeV},  \,
\, m_{1/2} \sim \frac{m_{3/2}^3}{M_{Pl}^2} \sim \mbox{TeV}
\end{equation}
so the gauginos and higgsinos are very naturally near the TeV
scale! While these estimates are rough, we have found a new link
between the TeV scale, here setting the dark matter mass, and the
GUT/Planck hierarchy. The scalar masses are more dependent on the
details of SUSY breaking. We next discuss a concrete model
implementing these ideas.

\subsection{Scherk-Schwarz models}

Models with $F_\phi \ll m_{3/2}$ arise naturally in the context of
no-scale models \cite{Lahanas:1986uc}, which can arise from
Scherk-Schwarz SUSY breaking in extra dimensions
\cite{Scherk:1979zr}, or equivalently from SUSY breaking by the
$F-$ component of a radion chiral superfield $T = r + \theta^2
F_T$ \cite{Kaplan:2001cg}. We follow the discussion of
\cite{Luty:2002hj}. Consider an extra dimension which is an
interval, and add a constant superpotential localized on one of
the boundaries, say the right boundary.  The tree-level low-energy
effective Lagrangian for $T$ and the chiral compensator $\phi = 1
+ \theta^2 F_\phi$ is of the form
\begin{equation}
L = \int d^4 \theta M_5^3 \phi^\dagger \phi (T + T^\dagger) + \int
d^2 \theta c M_5^3 \phi^3 + \mbox{h.c}
\end{equation}
leading to the scalar potential
\begin{equation}
V = M_5^3 \left[r |F_{\phi}|^2 + F_T^* F_\phi + 3 c F_\phi +
\mbox{h.c.} \right]
\end{equation}
The $F_T$ equation of motion fixes
\begin{equation}
F_\phi = 0
\end{equation}
while the $F_\phi$ equation of motion fixes
\begin{equation}
F_{T} = -3 c
\end{equation}
Therefore supersymmetry is broken, with vanishing tree-level
potential.  This is the famous ``no-scale" structure. The
goldstino is the fermionic component of $T$, which is eaten by the
gravitino, and the mass is
\begin{equation}
m_{3/2} \sim \frac{c}{r}
\end{equation}
while $F_\phi$ vanishes at this level. Hereafter we will assume $c
\sim 1$.

At 1-loop, a non-zero potential is generated--the gravitational
Casimir energy--which arises from a contribution to the K\"{a}hler
potential of the form
\begin{equation}
\int d^4 \theta \frac{1}{16 \pi^2} \frac{1}{(T + T^\dagger)^2}
\end{equation}
yielding
\begin{equation}
V_{grav} \sim -\frac{1}{16 \pi^2} \frac{1}{r^4}
\end{equation}
This tends to make the radius shrink. However, the addition of $N$
bulk hypermultiplets of mass $M$ can give rise to a repulsive
Casimir energy of the form
\begin{equation}
V_{Hyper} \sim + \frac{N}{16 \pi^2} \frac{1}{16 \pi^2} e^{-M r}
\end{equation}
Therefore there can be a minimum of the potential around $r \sim
M^{-1}$. The $F_T$ equation of motion now forces
\begin{equation}
F_{\phi} \sim \frac{1}{16 \pi^2} \frac{F_T}{M_5^3 r^4} \sim
\frac{1}{16 \pi^2} \frac{1}{M_5^3 r^4}
\end{equation}
so clearly $F_\phi \ll m_{3/2}$.

The value of the potential is negative at the minimum; in order to
cancel the cosmological constant, we have to have an additional
source of SUSY breaking on one of the branes, a superfield $X$
with $F_X \neq 0$ and
\begin{equation}
|F_X|^2 \sim \frac{1}{16 \pi^2} \frac{1}{r^4}
\end{equation}
To be concrete, suppose we have a chiral superfield $X$ localized
on the left boundary with a superpotential
\begin{equation}
W = \phi^3 m^2 X
\end{equation}
and a K\"{a}hler potential
\begin{equation}
K = \phi^\dagger \phi \left(X^\dagger X - \frac{(X^\dagger
X)^2}{M_5^2} + \mbox{higher powers of $X^\dagger X$} \right)
\end{equation}
This form of $W$ and $K$ can be guaranteed by an R-symmetry under
which $X$ has charge 2, although this is not necessary. If we
instead only assume that $X$ carries a spurious $U(1)$ charge -2
under which $m$ has charge +1, so that $X$ only appears in the
combinations $X^\dagger X$ and $m^2 X$, our conclusions are
unaltered.

Evidently $F_X \sim m^2$, in order to cancel the vacuum energy, we
have to choose
\begin{equation}
m^2 \sim \frac{1}{4 \pi} \frac{1}{r^2}
\end{equation}
The $(X^\dagger X)^2$ term in $K$ give the scalar component of $X$
a positive mass squared $\sim m^4/M_5^2$. The non-vanishing
$F_\phi$ gives a linear term to $X$ from the superpotential
coupling, so $X$ gets a vev. We then have a local minimum with
\begin{equation}
F_X \sim m^2, \, \, X \sim \frac{m^2}{M_5}
\end{equation}
The combination of the non-vanishing $F_X$ and $X$ also gives the
fermionic component of $X$, $\psi_X$, a mass of order
\begin{equation}
m_{\psi_X} \sim \frac{m^4}{M_5^3}
\end{equation}
From the $(X^\dagger X)^2$ part of $K$.

At this point we have  broken SUSY, stabilized the various moduli
and fine-tuned away the vacuum energy. The masses of all fields
can be expressed in terms of the 5D and 4D Planck scales by using
the usual flat space relationship $M_4^2 \sim M_5^3 r$, and we
find
\begin{equation}
m_{3/2} \sim \frac{M_5^3}{M_4^2}, \, \, m_{radion} \sim
\frac{M_5^6}{4 \pi M_4^5}, \, \, m_X \sim \frac{M_5^5}{4 \pi
M_4^4}, \, \, m_{\psi_X} \sim \frac{M_5^9}{16 \pi^2 M_4^8}
\end{equation}

The spectrum of the rest of the superpartners now depends on their
location in the bulk. We will assume that the SSM fields are
localized on the same brane as $X$, and therefore direct mediation
of SUSY breaking to the SSM scalars through operators of the form
\begin{equation}
\int d^4 \theta \frac{1}{M_5^2}X^\dagger X Q^\dagger Q
\end{equation}
are unsuppressed, leading to scalars masses of the same order as
$m_X$:
\begin{equation}
m_{S} \sim \frac{|F_X|}{M_5} \sim \frac{M_5^5}{4 \pi M_{4}^4}
\end{equation}

What about the gaugino masses? An irreducible source of $R$-
breaking is through the gravitino mass $m_{3/2}$. There is then a
finite 1-loop diagram, involving a propagator stretching between
the two boundaries of the extra dimension, giving a gaugino mass.
The magnitude can be estimated by drawing the diagram in the
low-energy theory cut-off off the scale $1/r$. The result is equal
for all gauginos and is
\begin{equation}
M^{\mathrm{grav}}_{3,2,1} \sim \frac{1}{16 \pi^2 M_4^2}
(\frac{1}{r})^3 \sim \frac{M_5^9}{16 \pi^2M_4^8}
\end{equation}
of the same order as $m_{\psi_X}$. Note that the operator
corresponding to this gaugino mass must is of the general form we
considered in the previous subsection,
\begin{equation}
\sim \int d^4 \theta \frac{c^\dagger}{M_5^3 (T + T^\dagger)^2}
W_\alpha W^\alpha
\end{equation}
This dominates over the anomaly mediated contribution by a
perturbative loop factor
\begin{equation}
M_{3,2,1}^{\mathrm{anom}} \sim \frac{g^2}{16 \pi^2} F_\phi \sim
\frac{g^2}{16 \pi^2} M^{\mathrm{grav}}
\end{equation}

In addition, we can have contact interactions on our brane which
can give rise to gaugino masses. The leading allowed operators are
of the form
\begin{equation}
\int d^2 \theta \frac{m^2 X}{M_5^3} WW; \, \, \int d^4 \theta
\frac{X^\dagger X}{M_5^3} WW + \mbox{h.c.}
\end{equation}
which generate gaugino masses of order
\begin{equation}
M_{i} \frac{|F_X|^2}{M_5^3} \sim M^{\mathrm{grav}}
\end{equation}
which are comparable to the gravitationally induced masses, and
can be different for $M_{3,2,1}$.

We now turn to $\mu B$ and $\mu$. As in the MSSM, something must
have suppressed the $M_5 H_u H_d$ term in the superpotential. This
could for instance be due to an accidental $R$- symmetry at the
level of the renormalizable couplings of the theory, under which
$H_{u,d}$ carry $R$- charge $0$. The leading allowed couplings are
then
\begin{equation}
\int d^2 \theta \frac{m^2 X}{M_5^2} H_u H_d; \, \, \int d^4 \theta
X^\dagger X H_u H_d; \, \, \int d^4 \theta \frac{m^2
X^\dagger}{M_5^3} H_u H_d
\end{equation}
these generate a $\mu B$ term of the appropriate size
\begin{equation}
\mu B \sim \frac{|F_X|^2}{M_5^3} \sim m_{S}^2
\end{equation}
As well as a $\mu$ term of the same order as the gaugino masses
\begin{equation}
\mu \sim M_i
\end{equation}

So, we have presented a simple model which breaks supersymmetry
with a stabilized extra dimension, producing an interesting
hierarchy of scales for the gravitino, scalars, gauginos and
Higgsinos of the theory:
\begin{equation}
m_{3/2} \sim \frac{\pi M_5^3}{M_4^2}; \, \, \,   m_{S} \sim
\frac{\pi M_5^5}{M_4^4}; \, \, \, M_i, \mu, m_{\psi_X} \sim
\frac{\pi M_5^9}{M_4^8}
\end{equation}
where in the above we have been more careful about the $2 \pi$
factors involved in the ratio of 5D and 4D Planck scales.

Let us get an idea for the scales involved. It is natural to use
this extra dimension to lower the higher dimensional Planck scale
down to the GUT scale, a la Horava-Witten \cite{HW}, $M_5 \sim M_G
\sim 3 \times 10^{16}$ GeV. Then we have
\begin{equation}
m_{3/2} \sim 10^{13} \mbox{GeV}; \, \, m_{S} \sim 10^9 \mbox{GeV};
\, \, m_{\mathrm{radion}} \sim 10^7 \mbox{GeV}; \, \, M,\mu \sim
100 \mbox{GeV}
\end{equation}
Note that even though there was no a priori reason for the
gauginos and Higgsinos to end up anywhere near the $\sim 100$ GeV
scale, they are in the right ball-park from this simple estimate.

We can also contemplate other sorts of theories of SUSY breaking
on the SSM brane where $R$ is more badly broken, such that $X$ and
$F_X$ are set by the same scale $X \sim \sqrt{F_X}$. In this case
the operator $X^\dagger X H_u H_d$ also generates a $\mu$ term of
the order of
\begin{equation}
\mu \sim \frac{X^* F_X}{M_5^2}
\end{equation}
which, if we make the reasonable assumption that $|X| \sim
\sqrt{|F_X|}$, gives rise to the estimate
\begin{equation}
\mu \sim (\frac{1}{4 \pi})^{3/2} \frac{M_5^7}{M_4^6}
\end{equation}
which is much bigger than $M^{\mathrm{grav}}$ or $M_i$. This is
also a potentially interesting scenario, since for the same
parameters as as above this $\mu$ is $\sim 100$ TeV. Again, if the
coefficient of the $H_u H_d$ are suppressed by a factor
$\epsilon$, this leads naturally to a large tan$\beta \sim
1/\epsilon$ which can be $\sim 10^{-2}$, so that $\mu$ can be
suppressed by a further tan$\beta$ factor to be near the $\sim$
TeV scale.

It should be clear that these theories are not particularly
engineered; we are breaking SUSY in one of the simplest possible
ways, and then simply following our nose to cancel the vacuum
energy and stabilize all the moduli. It certainly seems possible
that this sort of mechanism could be ``generic" within a large
neighborhood of the landscape.

Note that in addition to the SSM fermions, we have an additional
light fermion $\psi_X$ of comparable mass. Unlike the gravitino of
the usual SSM, however, this new particle does not give rise to
cosmological difficulties. If we assume that it is heavier than
the LSP, it can rapidly decay to it via the couplings giving rise
to the $\mu$ term; for instance the $(m^2/M_5^2) X H_u H_d$
operator gives rise to a decay width for $\psi_X$ of order
\begin{equation}
\Gamma \sim \frac{m^4}{M_5^4} m_{\psi_X}, \tau \sim 10^{-10}
\mbox{s}
\end{equation}
for $m_{\psi_X} \sim$ TeV. This decay happens well before
nucleosynthesis and poses no cosmological dangers.

\subsection{Non-SUGRA models with gauginos/Higgsinos $\sim 100$ GeV}

It is also possible that supersymmetry is so badly broken in the
gravitational sector of the theory that supergravity is not a good
low-energy approximation, but that nevertheless in the $M_{Pl} \to
\infty$ limit a globally supersymmetric field theory sector is
recovered. This can certainly be done within a consistent
effective theory, and amounts to working with a fixed cutoff $M_*$
which we will take to be $\sim M_{G}$, and including hard SUSY
breaking spurions suppressed by appropriate powers of
$(M_G/M_{Pl})$ in the effective theory. In this case, we can write
an effective action of the form e.g.

\begin{equation}
\int d^4 x \frac{1}{\epsilon} \sqrt{-g} M_G^2 R + {\cal L}_{SSM} +
\epsilon M_G^2 \left(\tilde{q}^* \tilde q  + \tilde l^* \tilde l +
h_{u,d}^* h_{u,d} + h_{u} h_d \right) + \cdots
\end{equation}
where the spurion $\epsilon$ is
\begin{equation}
\epsilon \sim \left(\frac{M_G}{M_{Pl}} \right)^2
\end{equation}
and we can also expect corrections to the dimensionless SUSY
couplings of $O(\epsilon)$. The scale $m_S$ is then naturally
\begin{equation}
m_S \sim \frac{M_G^2}{M_{Pl}} \sim 10^{13} - 10^{14} \mathrm{GeV}
\end{equation}

In such a model, there is no a priori need to break $R$ in order
to cancel the vacuum energy. It is amusing to contemplate the $R$-
symmetric limit of the SSM \cite{Hall:1990hq,Feng:1995dn}. The
most immediate problem is the massless gluino; although perhaps
this can be hidden in the QCD muck \cite{Farrar:1997ns}. However,
in the usual SSM, the $R$-symmetric limit also suffers from
problems in the electroweak sector: the charginos and one of the
neutralinos do get masses from electroweak symmetry breaking, but
they are too light: the sum of the chargino masses is smaller than
$2 m_W$, and the one of the neutralinos is degenerate with the
$Z$, both of which have been ruled out at LEP II. However in our
model with very high $m_S$, the low-energy Yukawa couplings
$\kappa_{u,d},\kappa^\prime_{u,d}$ are no longer forced to equal
the gauge couplings $g_2,g_1$ by SUSY, and in fact grow relative
to $g_{2,1}$ by RG scaling, which is also sensitive to the
possible presence of additional ($5 + \bar{5}$) multiplets beneath
the GUT scale. Thus the chargino/neutralino masses can become
heavier, and it may be possible to evade these direct detection
limits on the electroweak-inos even in the $R$-symmetric limit.

Nevertheless, since the massless gluino is so problematic
\cite{Janot:2003cr}, it is more reasonable to imagine that gaugino
and Higgsino masses are generated from some source of spontaneous
$R$- breaking in the low-energy theory. For instance, we can have
a hidden sector gauge group $G$ with fermions $\psi,\psi^c$ with
$R$-charge $-1$ (like the Higgsinos), and $R$- symmetric higher
dimension operators linking that sector to ours via
\begin{equation}
 \frac{\epsilon}{M_G^2} \psi^c \psi \lambda \lambda, \, \,
 \frac{\epsilon}{M_G^2} \bar{\psi^c} \bar{\psi} \psi_u \psi_d
\end{equation}
then if the $\psi, \psi^c$ condense at a scale $\Lambda$ we
generate gaugino/Higgsino masses
\begin{equation}
m_{1/2} \sim \mu \sim \epsilon \frac{\Lambda^3}{M_{G}^2}
\end{equation}
Note that there is no need to worry about $R$-axions associated
with the breaking of $R$; like the $\eta^\prime$, the would-be
Goldstone can get a mass from its anomaly with the hidden sector
gauge group.

In order to be able to make a prediction for these masses, we need
to know the particle content and $\Lambda$ scale for the hidden
sector. A natural assumption is that the hidden group is a unified
group like $SU(5)$ or $SO(10)$, and that the value of the coupling
at the GUT scale is equal to the SSM unified coupling, with
$\alpha_{GUT} = 1/33$ for scalars near $\sim 10^{13}$ GeV. If we
take SU(5) with a single $(5 + \bar{5})$ in the hidden sector,
then $m_{1/2}$ comes out to be too small, about $10^{-3}$ GeV.
However if we use SO(10), then $m_{1/2}$ is naturally near the
weak scale! For SO(10) with $N_T$ 10's, we find $m_{1/2} \sim \mu
\sim 1$ TeV for $N_T = 1$, and decreasing as $N_T$ increases, down
to $\sim 10$ GeV for $N_T = 8$ or equivalently a single adjoint of
SO(10). Once again, making a minimal set of assumptions, the
gaugino/Higgsino masses again end up ``accidentally" near the weak
scale.

\subsection{SUSY unification in non-SUSY natural theories}

For the readers who continue to pine for natural theories, it is
perhaps worth mentioning that it is possible to construct theories
with natural electroweak symmetry breaking without low-energy
SUSY, but with essentially supersymmetric gauge coupling
unification. Let us suppose that SUSY is broken and that now {\it
all} the scalars are heavy, but the gauginos and Higgsinos remain
light. Gauge coupling unification will still work well, but
another sector is needed for electroweak symmetry breaking. We can
imagine triggering this with strong dynamics as in technicolor or
composite Higgs models, or via AdS duals \cite{Maldacena:1997re}
of such theories. In order to preserve gauge coupling unification,
the EWSB sector must have a global $SU(5)$ or $SU(3)^3/Z_3$
symmetry, into which the SM is gauged in the usual way.

A sketch of an AdS representation of such an idea is as follows.
Consider a slice of AdS with SUSY broken in the bulk. There is an
$SU(3)^3/Z_3$ gauge symmetry in the bulk, broken by boundary
conditions to $SU(3)_c \times SU(2)_L \times U(1)_Y$ on the Planck
and IR branes. We have the gauginos and Higgsinos on the Planck
brane. Meanwhile, we have an elementary Higgs doublet on the IR
brane, so that in 4D CFT language we have a composite Higgs model
(we do this for ease of discussion; such models suffer from some
tunings and some extra model-building is needed to preserve
custodial $SU(2)$, by preserving the full $SU(2)_R$ on the IR
brane \cite{Agashe:2003zs}, but we ignore these details here). In
the bulk, we have 3 copies of the standard trinification
$(3,\bar{3},1) +$  cyclic multiplets, which contain (in SO(10)
language) the SM 16 together with an additional $10 + 1$. We can
decouple the extra $10 + 1$'s by marrying them off with elementary
fermions on the Planck brane, and the SM Yukawa couplings can be
generated by writing down Yukawa couplings between the appropriate
components of the bulk fermion and Higgs localized on the IR
brane; the fermion mass hierarchy can be generated by giving the
different generation fermions different bulk mass terms, which
localize them by varying amounts to the IR brane, this also avoids
large FCNC's \cite{Gherghetta:2003wm, Agashe:2003zs}. Note that
SU(3$)^3$ was chosen instead of SU(5) since in the SU(5) case,
bulk X/Y gauge boson exchange would give rise to unacceptably
large rates for proton decay.

It would be interesting to flesh out this construction. The gluino
lifetime will continue to be large, however, one would never know
whether the high-energy theory is really supersymmetric: while the
Higgsinos and gauginos survive to low energies, the scalar Higgs
does not, and there are therefore no dimensionless couplings that
bear an imprint of the high-scale supersymmetry, unlike the finely
tuned examples that have been the focus of this paper.

\section{Open Problems}
There are a number of computations of immediate phenomenological
importance in the high-scale supersymmetry scenario we have
outlined in this paper.
\begin{itemize}
\item{\bf Higgs Mass Prediction.} It is important do a full
analysis for the Higgs mass in this model; we have included part
of the full 1-loop running and the largest effects from threshold
corrections (most notably the 1-loop QCD correction to the top
mass), but a fully systematic analysis inclduing 2-loop running
and 1-loop threshold corrections is needed.

\item {\bf Gluino Phenomenology.} The gluino is perhaps the most
important particle of this framework, as its lifetime is
 a direct probe of the SUSY breaking scale. Moreover,
the LHC can be a gluino factory, and therefore an ideal place to
study its properties.
 Understanding the gluino energy loss and
looking for charged tracks, intermittent tracks, displaced vertices and,
especially, delayed off-time decays, can help us measure the
 gluino lifetime and open a window into scales of supersymmetry breaking as high as $10^{13}$ GeV.

\item{\bf Higgsino-Gaugino Phenomenology.} Winos and higgsinos
will be produced through Drell-Yan --and not, as typical in the
SSM, by squark/gluino production followed by a cascade of decays
down to the LSP. It is important to investigate how accurately we
can measure the Higgs-Higgsino-Gaugino Yukawa couplings at
colliders, possibly at the LHC but more likely at a linear
collider. These measurements extrapolated to high energies can
give striking evidence for high-scale SUSY.

\item{\bf Two-Loop Corrections to Unification.} As we have seen,
our 1-loop prediction for $\alpha_3(M_Z)$ is somewhat lower than
in the SSM; it is important to perform the full two loop analysis,
as this will likely push up our $\alpha_3(M_Z)$ into better
agreement with experiment.

\item{\bf Dark Matter Detection and Abundance.} Because of the
absence of scalars, the collision and annihilation cross sections
of the lightest neutralino depend on fewer (than in the SSM)
parameters. So, a proper computation of these processes is
important, as it can help pin down the interesting parameter
ranges in our low-energy theory. It may also predict more
precisely the DM detection cross sections.

\item{\bf Gluino Cosmology.} As we argued, the gluinos can undergo
a second stage of annihilation around the QCD phase transition
that further depletes their abundance relative to the standard
perturbative freeze-out calculation. It is important to understand
this in detail, as this can determine the allowed ranges for the
gluino lifetime, and thereby affect the allowed masses for the
heavy scalars.

\end{itemize}

\section{Travel Guide to a Finely-Tuned World}

Although the cosmological constant problem casts a giant shadow on
the principle of naturalness, the prevailing view has been that
the LHC will reveal a natural theory for electroweak symmetry
breaking, and that gauge coupling unification favors this to be
\emph{low-energy} SUSY, despite its nagging problems and the
accompanying epicyclic model-building needed to address them.

Here we have outlined an alternate viewpoint, where the usual
problems of SUSY vanish, unification is evidence for
\emph{high-energy} SUSY, and where accelerators can convincingly
demonstrate the presence of fine tuning in the electroweak sector.

The first sign of this proposal at the LHC should be the Higgs, in
the mass range of $\sim 120-150$ GeV. No other scalar should be
present, since it would indicate a second, needless, fine-tuning.
Next will be the gluino, whose long lifetime will be crucial
evidence that the scale of supersymmetry breaking is too large for
the hierarchy problem, and a fine-tuning is at work. A measurement
of the gluino lifetime can yield an estimate for the large SUSY
breaking scale $m_S$.
Next will come the electroweak gauginos and higgsinos, whose
presence will complete the picture, and give supporting evidence
that the colored octets of the previous sentence are indeed the
gluinos. Further precise measurements of the
gaugino-higgsino-higgs couplings, presumably at a linear collider,
will accurately determine $m_S$ and provide several unambiguous
quantitative cross-checks for high-scale supersymmetry.

If this scenario is confirmed experimentally, it will be a
striking blow against naturalness, providing sharp evidence for
the existence of supersymmetry in nature, as may have been
expected for a consistent UV theory of gravity, but not at low
enough scales to solve either the hierarchy or cosmological
constant problems. This will strongly point to a very different
set of ideas to explain these fine-tunings -- such as the
``galactic" and ``atomic" principles, selecting the vacuum of our
finely tuned world from a small neighborhood in a landscape of
vacua. This may be the closest we will ever come to direct
experimental evidence for this vast landscape.

\section{Acknowledgments}
It is a pleasure to thank Spencer Chang, Hsin-Chia Cheng, Paolo
Creminelli, Glennys Farrar, Gregory Gabadadze, JoAnne Hewett,
Gordy Kane, Lubo\v{s} Motl, Aaron Pierce, Tom Rizzo, Eva
Silverstein, Jay Wacker, Neal Weiner and Matias Zaldariaga for
valuable discussions. Special thanks to Shamit Kachru, Andrei
Linde, and Lenny Susskind for guiding us through the landscape,
Gia Dvali and Markus Luty for important comments on SUSY breaking,
and Matt Strassler for pointing out many novel features of the
long-lived gluino's collider phenomenology. Thanks also to Jay
Wacker and Aaron Pierce for pointing out omissions and typos in
the Higgs quartic coupling RGE's given in the first version of
this paper. SD would like to thank the Harvard theory group for
its hospitality. The work of NAH is supported by the DOE grant
DE-FG02-91ER40654 and the David and Lucille Packard foundation. SD
is supported by NSF Grant 0244728.


\providecommand{\href}[2]{#2}\begingroup\raggedright\endgroup

\end{document}